\documentclass[aps,twocolumn,prd,showpacs,nofootinbib]{revtex4}
\usepackage{amsmath}
\usepackage{graphicx}
\usepackage{dcolumn}
\usepackage{bm}
\usepackage{amssymb}
\usepackage{latexsym}

\def\be{\begin{equation}}
\def\ee{\end{equation}}
\def\ba{\begin{eqnarray}}
\def\ea{\end{eqnarray}}

\begin{document}

\title{Primordial
Perturbations During a Slow Expansion}

\author{Yun-Song Piao}

\affiliation{College of Physical Sciences, Graduate School of
Chinese Academy of Sciences, Beijing 100049, China}


\begin{abstract}

Recently, it has been shown that a slow expansion, which is
asymptotically a static state in infinite past and may be
described as an evolution with $\epsilon\ll -1$, of early universe
may lead to the generation of primordial perturbation responsible
for the structure formation of observable universe. However, its
feasibility depends on whether the growing mode of Bardeen
potential before phase transition can be inherited by the constant
mode of curvature perturbation after phase transition. In this
note, we phenomenally regard this slow expansion as that driven by
multi NEC violating scalar fields. We calculate the curvature
perturbation induced by the entropy perturbation before phase
transition, and find that the spectrum is naturally scale
invariant with a slight red tilt. The result has an interesting
similarity to that of slow roll inflation.

\end{abstract}

\pacs{98.80.Cq}

\maketitle

The results of recent observations \cite{WMAP} are consistent with
an adiabatic and nearly scale invariant spectrum of primordial
perturbations, as predicted by the simplest models of inflation.
The inflation stage is supposed to have taken place at the earlier
moments of the universe \cite{Guth, LAS, S80}, which
superluminally stretched a tiny patch to become our observable
universe today. During the inflation the quantum fluctuations in
the horizon will be able to leave the horizon and become the
primordial perturbations responsible for the formation of
cosmological structure
\cite{BST83, MC}. This is one of remarkable successes of
inflation,
However, this success is also shared by an expansion with the null
energy condition (NEC) violation, which corresponds to e.g. the
phantom inflation \cite{PZ, PZ1, GJ, BFM, ABV}, see also Ref.
\cite{DHC} for comments. The reason is that the inflation can be
generally regarded as an accelerated or superaccelerated stage,
and so may defined as an epoch when the comoving Hubble length
decreases, which occurs equally during the NEC violating
expansion. It is the shrinking of this comoving Hubble length that
leads the causal generation of primordial perturbations.

The primordial perturbation generated during the NEC violating
evolution has been studied in Ref. \cite{PZ, PZ1}. In Ref.
\cite{PZ}, it was firstly noticed that there is an interesting
limit case in which $\epsilon\ll -1$, where $\epsilon$ is defined
as $-{\dot h}/h^2$ and $h$ is the Hubble parameter, which
corresponds to that the scale factor grows very
slowly but the Hubble length rapidly shrinks. 
During the slow expansion the primordial perturbation can be
generated, see Fig.1. The end of slow expanding phase may be
regarded as a reheating process or phase transition that the
fields dominating the background decay into usual radiation, which
then will be followed by a FRW evolution of standard cosmology. We
found that the spectrum of Bardeen potential $\Phi$ before the
transition is dominated by an increasing mode and is nearly scale
invariant \cite{PZ}. Though during this period the spectrum of
comoving curvature perturbation $\xi$ is strong blue, if the
growing mode of spectrum of Bardeen potential before the
transition may be inherited by the constant mode of $\xi$ after
the transition, which is similar to the case \cite{DurrerV, CDC,
GKST} of the ekpyrotic/cyclic scenario \cite{KOST, STS}, the
spectrum of resulting adiabatic fluctuations appearing at late
time will be scale invariant. However, it is obvious that the
result is severely dependent of whether this inheriting can occur,
which is actually determined by the physics at the epoch of phase
transition. Thus there is generally an uncertainty. In the simple
and conventional scenario it seems that the growing mode of $\Phi$
can hardly be matched to the constant model after the transition
\cite{BF, Lyth, Hwang}, which has been shown by some numerical
studies \cite{C, AW, BV}. Further, it has been illuminated
\cite{DurrerV, Bozza} that whether the final spectrum is that of
the growing mode before the transition depends on whether there is
a direct relation between the comoving pressure perturbation and
$\Phi$ in the energy momentum tensor, in which the new physics
mastering the transition might be encoded.
Thus with these points it seems that though whether the nearly
scale invariant primordial perturbation may be generated during a
slow expansion of early universe is still open, the possibility
remains.

The slow expansion with $\epsilon\ll -1$ may have some interesting
applications in cosmology. For example, the semiclassical studies
of island universe model, in which initially the universe is in a
cosmological constant sea, then the local quantum fluctuations
with the NEC violation will create some islands with matter and
radiation, which under certain conditions might correspond to our
observable universe \cite{DV, Piao0506, Piao0512}. Thus with the
debate whether the scale invariant spectrum of curvature
perturbation may be obtained during such a slow expansion, the
study of relevant issues is quite interesting.
Note that in Ref. \cite{PZ}, we adopt the work hypothesis that the
NEC violating phase with $\epsilon\ll -1$ is implemented by a
scale field with the NEC violation, in which the scalar field has
a reverse sign in its dynamical terms. Thus it may be conceivable
that our hypothesis and simplified operation in the calculations
of primordial perturbation spectrum might have missed what. In
this paper, we will study a slight nontrivial case, in which the
slow expansion with $\epsilon\ll -1$ is simulated phenomenally by
that driven by multi scalar fields with the reverse sign in their
dynamical terms. We find that the spectrum of entropy perturbation
is scale invariant with a slight red tilt. The curvature
perturbation under certain conditions may be induced by the
entropy perturbation, and thus may has the same spectral index
with the entropy perturbation. We show that the spectrum and
amplitude of curvature perturbation induced by the entropy
perturbation at the end epoch of the NEC violating phase can be
related to those of inflation by a dual invariance.

Firstly, let us briefly review the results of Ref. \cite{PZ}. For
a slow expansion with the NEC violation, the evolution of scale
factor $a(t)$ may be simply taken as \be a(t) \sim {1\over
(-t)^{n}}\sim (-\eta)^{-{n\over n+1}},\label{at}\ee where $n\ll 1$
is a positive constant. When $t$ is initially from $-\infty$ to
$0_-$, it corresponds to a slow expansion. The Hubble parameter is
\be h= {n\over (-t)}, ~~~~~ {\dot h}= {n\over (-t)^2}, \label{h}
\ee thus $\epsilon = -{1/ n}\ll -1$. The $\epsilon$ can be
rewritten as $\epsilon \simeq {1\over h \Delta t}{\Delta h\over
h}$, thus in some sense $\epsilon$ actually describes the change
of $h$ in unit of Hubble time and depicts the abrupt degree of
background background. From Eq.(\ref{h}), during the slow
expansion, though the scale factor is hardly changed, the Hubble
parameter will rapidly increase, which means an abrupt change of
background \footnote{When $t$ approaches negative infinity, we
have $h\rightarrow 0$, which means that the universe is
asymptotically a static state in infinite past. This in some sense
is similar to an emergent universe studied in Ref. \cite{EM}, see
also \cite{EMT, MTLE}, in which the initial static state is
constructed by introducing a positive curvature. However, here it
corresponds to be implemented by using a scalar field with the NEC
violation, in which the initial kinetic energy of scalar field
just approximately sets off its potential energy.}. In Ref.
\cite{PZ}, it was showed that when the slow expansion is
implemented by a scalar field with a reverse sign in its dynamical
term, the spectral index of Bardeen potential $\Phi$ is given by
\be n_{\Phi}-1\simeq 2n, \label{nphi}\ee which is nearly scale
invariant with a slightly blue tilt. When the optimistic matching
of the growing mode of $\Phi$ before the phase transition to the
constant model of $\xi$ after the phase transition can be made,
the amplitude after the end of slow expanding phase is given by
\cite{Piao0506} \be {\cal P}_{(\Phi\rightarrow \xi)}\cong {1\over
n}\cdot \left({h_e\over 2\pi}\right)^2 ,\label{Phik}\ee where
$G=1$ has been set and the subscript `e' denotes the end epoch of
slow expansion.

\begin{figure}[t]
\begin{center}
\includegraphics[width=8cm]{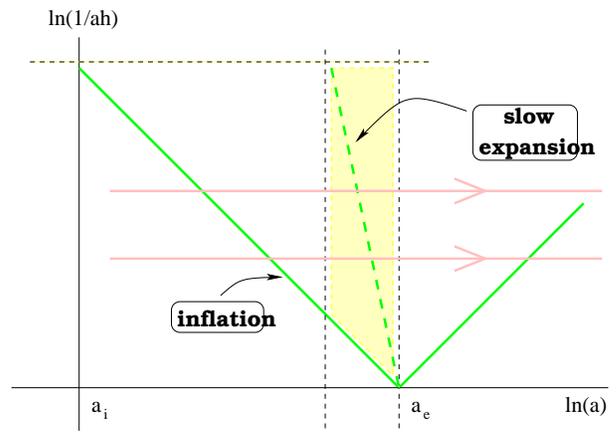}
\caption{The evolutive figure of $\ln{(1/ah)}$ with respect to the
scale factor $\ln{a}$ during the slow expansion with $\epsilon\ll
-1$, which is compared with that of slow roll inflation. The
details that the figure is plotted can be seen in Ref.
\cite{Piao0609}, in which $a_e$ denotes that at the end of slow
expanding phase. The red lines are the perturbation modes with
some wavenumber $k$. }
\end{center}
\end{figure}

Then let us see what occurs when the slow expansion is simulated
phenomenally by that driven by two or more NEC violating scalar
fields with the reverse sign in their dynamical terms. In this
case there is not only the curvature perturbation but the entropy
perturbation. No loose generality, we will study the case with two
scalar fields $\varphi_1$ and $\varphi_2$. Note that there exists
a scale solution in which ${\dot \varphi}_1/{\dot \varphi}_2$ is a
constant. In this case, the background values of all relevant
quantities of fields can be determined simply. We may write ${\dot
\varphi}_1$ and ${\dot \varphi}_2$ as \be {\dot \varphi_1} =
\sqrt{n_1\over 4\pi } {1\over (-t)},\,\,\,\,\, {\dot \varphi_2} =
\sqrt{n_2\over 4\pi } {1\over (-t)},\label{dotphi}\ee where both
$n_1$ and $n_2$ are positive constants. When $n_1+n_2=n$ is taken,
where $n$ is given by Eq.(\ref{at}), we may have \be
V(\varphi_1,\varphi_2) = {n(3n +1) \over 8\pi }{1\over (-t)^2},
\label{vphi}\ee which can be obtained by combining Eqs. (\ref{h})
and (\ref{dotphi}) and Friedmann equation. We see that for
arbitrary value $n>0$, $V(\varphi)$ is always positive, which is
different from that of the usual scalar field, in which when
$n<{1/ 3}$, the potential must be negative \cite{HW, GPZ0304}. The
reason is that here what we use is the scalar fields with the
reverse sign in their dynamical terms. Integrating (\ref{dotphi}),
and substituting the result obtained into (\ref{vphi}), we can
split the effective potential (\ref{vphi}) into two parts for
$\varphi_1$ and $\varphi_2$, respectively, \be V(\varphi_1)=
{n_1(3n+1)\over 8\pi } \exp{\left(-\sqrt{16\pi \over n_1}
\varphi_1\right)},\label{vv1}\ee \be V(\varphi_2)= {n_2(3n+1)\over
8\pi} \exp{\left(-\sqrt{16\pi \over n_2}
\varphi_2\right)}.\label{vv2}\ee Thus both fields are decoupled.
Note that $n\ll 1$, thus $n_1, n_2\ll 1$, Eqs.(\ref{vv1}) and
(\ref{vv2}) suggests that the potential of both $\varphi_1$ and
$\varphi_2$ are very steep. During the slow expansion, they will
climb up along their potentials, which is determined by the
property of the NEC violating field, e.g. \cite{ST, GPZ}. In this
case, it may be showed that this scale solution is an attractor,
e.g. see Ref. \cite{GPCZ}.

Before calculating the primordial perturbation, we need to
decompose these two fields into the field $\varphi$ along the
field trajectory, and the field $s$ orthogonal to the trajectory
by making a rotation in the field space as follows \be
\varphi={\sqrt{n_1}\varphi_1+\sqrt{n_2}\varphi_2\over
\sqrt{n}},\,\,\,\,s={\sqrt{n_2}\varphi_1-\sqrt{n_1}\varphi_2\over
\sqrt{n}}, \label{phis}\ee as has been done in Ref. \cite{GWBM}.
In this case, the potential (\ref{vphi}), which is the sum of
Eqs.(\ref{vv1}) and (\ref{vv2}), can be equivalently rewritten as
$U(s)\exp{\left(-\sqrt{16\pi \over n}\varphi\right)} $, where \ba
U(s) & = & {n_1(3n+1)\over
8\pi} \exp{\left(-\sqrt{16n_2\pi \over n_1 n} s\right)}\nonumber\\
& + & {n_2(3n+1)\over 8\pi} \exp{\left(-\sqrt{16n_1\pi  \over n_2
n} s\right)}\ea is the potential of $s$ field, whose effective
mass is given by $\mu^2(s)= U^{\prime\prime}(s)$. Thus we have \ba
{\mu^2(s)\over h^2}&=& {2n_2(3n+1)\over nh^2}
\exp{\left(-\sqrt{16n_2\pi  \over n_1 n} s\right)}\nonumber\\ & +
& {2n_1(3n+1)\over nh^2}
\exp{\left(-\sqrt{16n_1\pi\over n_2 n} s\right)}\nonumber\\
& \equiv & {2(3n+1)\over n^2}, \label{mus}\ea where Eqs.(\ref{h})
and (\ref{phis}) have been used. The result is not dependent of
$n_1$ and $n_2$, but only the background parameter $n$.

When this rotation is done, the perturbations will also generally
decomposed into two parts, one is the curvature perturbation
induced by the fluctuation of $\varphi$ field, and the other is
the entropy perturbation induced by the fluctuation of $s$ field.
In linear order, as long as the background trajectory remains
straight in field space, the entropy perturbation must be
decoupled from the curvature perturbation, which actually can be
seen in Ref. \cite{GWBM}. For the slow expansion, when the entropy
perturbation is decoupled from the curvature perturbation, the
calculation of curvature perturbation is the same as that of
single NEC violating field in Ref. \cite{PZ}, in which only when
the growing mode of $\Phi$ before the phase transition may be
inherited by the constant model of $\xi$ after the phase
transition, the spectrum is scale invariant, see Eqs.(\ref{nphi})
and (\ref{Phik}), or the spectrum will be strong blue, whose
amplitude is negligible on large scale. While the entropy
perturbation $\delta s$ may be calculated in the following way. In
the momentum space, the equation of entropy perturbation can be
given by \be v_k^{\prime\prime}+(k^2-f(\eta))v_k =0,
\label{uki}\ee where $\delta s_k\equiv v_k/a$ has been defined and
the prime denotes the derivative with respect to the conformal
time, and $f(\eta)$ is generally given by
\ba f(\eta) &\equiv & {a^{\prime\prime}\over a}+\mu^2(s)a^2\nonumber\\
&\simeq & {2+3n\over \eta^2},\,\,\,\,{\rm for}\,\, n\simeq 0_-,
\label{f}\ea where Eqs.(\ref{mus}) and $1/\eta= (1+1/n) ah$ has
been used. Note that the right side of the first line in
Eq.(\ref{f}) is the plus between two terms, but not minus as
usual, which is actually the result using the fields with the
reverse sign in their dynamical terms.

The solutions of Eq.(\ref{uki}) are Hankel functions. In the
regime $k\eta \rightarrow \infty $, all interesting modes are very
deep inside the horizon of the slow expanding phase, thus
Eq.(\ref{uki}) can be reduced to the equation of a simple harmonic
oscillator, in which $v_k \sim e^{-ik\eta} /(2k)^{1/2}$, which in
some sense suggests that the initial condition can be taken as
usual Minkowski vacuum. In the superhorizon scale, i.e.
$k\eta\rightarrow 0$, in which the modes become unstable and grow,
the expansion of Hankel functions to the leading term of $k$ gives
\be v_k\simeq {1\over \sqrt{2k}}(-k\eta)^{{1\over 2}-v}
,\label{uk}\ee where $v\simeq 3/2+n$, which may be deduced from
Eq.(\ref{f}), and here the phase factor and the constant with
order one have been neglected. During the slow expansion the
change of Hubble parameter $h$ is quite abrupt, as has been
pointed out. Thus it may be expected that the perturbation
amplitude of $v_k$ will continue to change after the corresponding
perturbations leaving the horizon, up to the end of the slow
expanding phase. This can also be explained as follows. When
$k\eta\rightarrow 0$, which corresponds to the super horizon
scale, we have $ v_k^{\prime\prime}-(2+3n)v_k/\eta^2 \simeq 0$.
This equation has one growing solution and one decay solution. The
growing solution is given by $ v_k\sim a^{1/n}$, where
Eq.(\ref{at}) has been used. The scale factor $a$ is nearly
unchanged, but since $n\simeq 0$, the change of $v_k$ has to be
significant, thus generally one can not obtain that the $|\delta
s_k |=|v_k/a|\sim a^{1/n}$ is constant, which actually occurs only
for the slow roll inflation in which approximately $n\rightarrow
\infty$. The details can also be seen in Ref. \cite{Piao0608}, in
which the spectrum of a test scalar field not affecting the
evolution of background was calculated, which in some sense
corresponds to the case of $n_2=0$ here. This suggests that in
principle we should take the value of $v_k$ at the time when the
slow expansion ends to calculate the amplitude of perturbations.
Thus the perturbation spectrum is \be 
k^{3/2} |{v_k(\eta_e)\over a}|
\sim k^{3/2-v}, \label{ps}\ee which suggests that the spectrum
index is given by $n_s-1\equiv 3-2v$. This leads to \be n_s-1
\simeq -2n, \label{ns}\ee which means that during the slow
expansion the spectrum of entropy perturbation is nearly scale
invariant with a slightly red tilt, since $n\simeq 0_+$. This
result is only determined by the evolution of background during
the slow expansion, but not dependent of other details.

We can see that if $|\epsilon|\sim 10^2$, the spectrum of entropy
perturbation may be very naturally matched to recent observation
\cite{WMAP}, since $n \equiv 1/|\epsilon |\sim 0.01$. Thus it may
be interesting to consider how these entropy perturbations can be
responsible for the structure formation of observable universe. To
do so, we need to make the curvature perturbation at late time
have an opportunity to inherit the characters of entropy
perturbation generated during the slow expansion. This can be
accomplished by noting that the entropy perturbation sources the
curvature perturbation \be |{\dot \xi}|\simeq {h{\dot\theta}\over
{\dot\varphi}}\delta s \label{dotxi}\ee on large scale
\cite{GWBM}, where $\theta\equiv {\rm arctg}\sqrt{n_2\over n_1}$
dipicts the motion trajectory of fields in field space of
$\varphi_1$ and $\varphi_2$, see Eq.(\ref{dotphi}). When $\theta$
is a constant, the trajectory is a straight line. In this case,
${\dot\theta}=0$, thus the entropy perturbation is decoupled from
the curvature perturbation, which also assures the validation of
Eq.(\ref{uki}), or there will some terms such as $\sim {\dot
\theta}^2$ and $\sim {\dot \theta}\Phi$. However, if there is a
sharp change of field trajectory, ${\dot\theta}$ must be not equal
to $0$, in this case $\dot \xi$ will inevitably obtain a
corresponding change induced by $\delta s$ by Eq.(\ref{dotxi}), as
has been pointed out and applied in ekpyrotic model \cite{LMTS,
BKO}, see also earlier Refs. \cite{NR, DFB} and recent studies
\cite{KW} on the ekpyrotic collapse with multiple fields. It may
be expected that at the end epoch of slow expanding phase the
scale solution will generally be broken, which also actually may
be constructed by modifying the potential of fields around the
end. For example, around the end epoch, instead of being steep,
the potential of one of fields will has a maximum or a plateau,
which will lead to the rapid stopping of up climbing of
corresponding field, while the up climbing of another field
remains, note that here the motion of field is mainly managed by
its potential, see e.g. Refs. \cite{ST, GPZ}. In this case, the
entropy perturbation will be able to source the curvature
perturbation.

We assume, for a brief analysis, that at split second before the
end of slow expanding phase the motion of $\varphi_2$ rapidly
stops while $\varphi_1$ remains, and then the phase transition
occurs and the universe quickly thermalizes into a radiation phase
and evolve with standard FRW cosmology. Following Ref. \cite{LMTS,
BKO}, this corresponds to a sharp change from initial fixed value
$\theta_{*}={\rm arctg}\sqrt{n_2\over n_1}$ to $\theta\simeq 0$.
It is this change that leads that $\xi$ acquires a jump induced by
the entropy perturbation and thus inherits the nearly scale
invariant spectrum of the entropy perturbation. In the rapid
transition approximation, one has obtained \be |\xi | \simeq
\theta_{*} {h_e\over {\dot\varphi}}\delta s
\simeq {h_e\over {\dot\varphi}}\delta s, \label{xi}\ee where the
constant factor with order one have been neglected. From
Eq.(\ref{ps}), the amplitude of entropy perturbation can be
calculated at the time when the slow expansion ends and given by
\be k^{3/2} |{v_k(\eta_e)\over a}|
\simeq {1\over n}\cdot\left({h_e\over 2\pi}\right), \label{psa}\ee
where $n\ll 1$ has been used. The calculations are similar to that
done in Ref. \cite{Piao0608}. The prefactor $1/n$ is from the
relation $1/\eta_e= (1+1/n) a_eh_e$, which corresponds the $g$
factor introduced and discussed in Ref. \cite{Piao0608}. Note that
$h^2/{\dot\varphi}^2\simeq -1/\epsilon =n $, thus we have the
amplitude of curvature perturbation \be {\cal P}_{(s\rightarrow
\xi)}\cong \left({h\over {\dot\varphi}}\right)^2\cdot k^3
\left|{v_k(\eta_e)\over a}\right|^2 \simeq {1\over
n}\cdot\left({h_e\over 2\pi}\right)^2. \label{pxi}\ee We can see
that this result is the same as Eq.(\ref{Phik}) in form, only up
to a numerical factor with unite order. Thus for the slow
expanding phase with $n\ll 1$ or equally $\epsilon \ll -1$, it
seems that whether induced by the increasing mode of Bardeen
potential, or by the entropy perturbation before the phase
transition, the resulting curvature perturbations after phase
transition is nearly scale invariant with the amplitude described
by the same equation at least in form. Though this dose not means
that the scalar spectrum of slow expansion must be scale
invariant, it seems at least that there are some convinced
possibilities that it may be.


The amplitude of usual slow roll inflation models with
$\epsilon\simeq 0$ may generally written as \be {\cal
P}_{\xi}\cong {1\over \epsilon}\cdot\left({h\over 2\pi}\right)^2.
\label{inpxi}\ee For $\epsilon$ approximately being a constant,
which corresponds to the case of scale solution in which the
inflation is driven by the scalar field with an exponent
potential, the spectra index is given by $n_s-1\simeq -2\epsilon$.
Thus we can see that they can be related to Eqs.(\ref{pxi}) and
(\ref{ns}) by making a replacement $|\epsilon |\rightarrow n$.
During the slow expansion the spectral index of curvature
perturbation induced by the increasing mode of Bardeen potential
is given in Eq.(\ref{nphi}), which is slightly blue tilt. Thus for
this case, it is also suitable for above replacement. 
This replacement may be regarded as a dual transformation between
their background evolutions, i.e. between the nearly exponent
expansion with $n\rightarrow \infty$, since here $|\epsilon
|\equiv 1/n$, and the slow expansion $n\simeq 0$. This extends the
studies on the dualities of the primordial density perturbation in
Refs. \cite{KST, BST, Piao, Lid} and see also recent Ref.
\cite{CZ} \footnote{see also Refs. \cite{CL, DSS, Calca, Cai0609}
for the discussions on the dualities of scale factor.}.

In summary, we phenomenally regard the slow expansion with
$\epsilon\ll -1$ as that driven by multi NEC violating fields. We
calculate the curvature perturbation induced by the entropy
perturbation before phase transition, and find that the spectrum
is naturally scale invariant with a slight red tilt, which may fit
recent observations well. This result to some extent highlights
the fact again that a slow expansion, which may be described as an
evolution with $\epsilon\ll -1$ and might be asymptotically a
static state in infinite past, before usual FRW evolution may be
feasible for seeding of the primordial perturbation responsible
for the structure formation of observable universe. Though
we still lack of understandings on some aspects of the phenomena
with the NEC violation, which might be quantum, we think that this
work, regarded as a semiclassical and effective description
\cite{KO}, might in some sense have captured some basic
ingredients of the NEC violating evolution of early universe,
which may be interesting and significant for many applications in
cosmology.




\textbf{Acknowledgments} The author thanks David Coule for
discussions. This work is supported in part by NNSFC under Grant
No: 10405029, in part by the Scientific Research Fund of
GUCAS(NO.055101BM03), in part by CAS under Grant No: KJCX3-SYW-N2.

\end{document}